\documentclass{PoS}

\title{Red halos and extragalactic background light}

\ShortTitle{Red halos and extragalactic background light}

\author{\speaker{Erik Zackrisson}\\
        Department of Astronomy, Stockholm University\\
        E-mail: \email{ez@astro.su.se}}

\author{Genoveva Micheva\\
Department of Astronomy, Stockholm University\\
        E-mail: \email{genoveva@astro.su.se}}
        
\abstract{Deep surface photometry of disk galaxies at optical and near-IR wavelengths have revealed faint halos with colours much too red to be reconciled with the resolved stellar populations detected in the halos of the Milky Way or M31. Alternative scenarios involving high metallicities, nebular emission or large amounts of dust in these halos are also disfavoured. Here, we argue that extinction of the optical extragalactic background light in the halos of these galaxies may possibly be responsible for the reported colour anomalies. We also discuss how an independent measurement of the optical extragalactic background light might be accomplished by combining direct star counts with surface photometry for a single nearby galaxy.}

\FullConference{Cosmic Radiation Fields: Sources in the early Universe\\
		 November 9-12, 2010\\
		 Desy, Germany}

\begin{document}

\section{Introduction}
The stellar halos of galaxies contain the fossil record of galaxy assembly, and observations of resolved halo stars in the Milky Way, Andromeda and other nearby galaxies have provided a wealth of information on such structures (e.g. \cite{Mouhcine et al.,Ibata et al.,Helmi et al.,Rejkuba et al.}). However, stellar halos can also be studied through surface photometry, i.e. observations of the integrated light from large numbers of unresolved stars within each system. 

At the current time, the results from these two techniques are difficult to completely reconcile. A number of attempts to study the halos of edge-on disk galaxies through optical and near-infrared surface photometry have revealed halo colours that are too red to be consistent with the halo populations captured through the study of resolved stars (e.g. \cite{Sackett et al.,Lequeux et al. a,Lequeux et al. b, Rudy et al.,James & Casali,Lequeux et al. c,Zibetti et al.,Zibetti & Ferguson,Bergvall et al.}). The most recent observations indicate that this red excess turns up at extremely faint surface brightness levels ($\mu_i\approx 27$--29 mag arcsec$^{-2}$ \cite{Zibetti et al.,Bergvall et al.}) and at wavelengths around the $i$-band ($\approx 7600$ \AA). Older measurements \cite{Rudy et al.,James & Casali} suggest that it may also continue into the near-infrared (12000--22000 \AA). 

It has been suggested that these `red halos' could be due to mundane observational effects like instrumental scattering \cite{de Jong} or flux residuals left over from an incomplete sky subtraction \cite{Zheng et al.,Jablonka et al.}. In some of the claimed detections, this may well be the case. However, Bergvall et al. \cite{Bergvall et al.} correct for instrumental scattering (point spread function) effects and are still left with a pronounced red excess in their study of halos around stacked, edge-on low surface brightness disks. They also show that while the halos of these disks have extremely red colours, the disks themselves do not display any anomalous colours at the corresponding surface brightness levels. This strongly argues against a sky flux residual as the main culprit, since this would result in strange colours in both components.

In a recent paper \cite{Zackrisson et al. a}, we argue that the colour anomalies could be due to extinction of the optical extragalactic background light (EBL).  While most of the sky flux originates from regions between the telescope and the halos targeted by surface photometry observations, a small fraction -- the EBL -- comes from behind these halos. Provided that the existing direct measurements of the optical EBL \cite{Bernstein} are not substantially affected by systematic problems, most of this light appears to be diffuse (i.e. unresolved with all existing instruments). Unlike the other components of the night sky flux, the EBL can be subject to extinction by dust present in the target halos. This invalidates an implicit assumption used in all current surface photometry measurements, namely that the sky flux and the flux from the target objects are unrelated. This extinction of the EBL may lead to a slight depression of the overall sky flux across the target halos, which is neglected when the sky flux level is estimated well away from the target objects. At faint surface brightness levels, even very small amounts of extinction would suffice. The net effect is a systematic, wavelength-dependent oversubtraction of sky flux, with spurious colour gradients in the outskirts of extended targets as the likely result. The resulting colours are sensitive both to the dust opacity and the intrinsic galaxy colours at the relevant isophotes. As argued by Zackrisson et al. \cite{Zackrisson et al. c} and Bergvall et al. \cite{Bergvall et al.}, this could explain why anomalous colours turn up only in the halo region and not at comparable surface brightness levels along disks -- since both the dust content and the intrinsic population colours may differ between these two components. To produce the required effect, no more than $A(g)\approx 0.01$--0.1 mags of optical extinction would be required in the halo, which is consistent with the typical halo opacities inferred from the reddening of background light sources \cite{Zaritsky,Menard et al.,McGee & Balogh}.

\section{Signatures of EBL extinction}
\begin{figure}
\centering
\includegraphics[width=.6\textwidth]{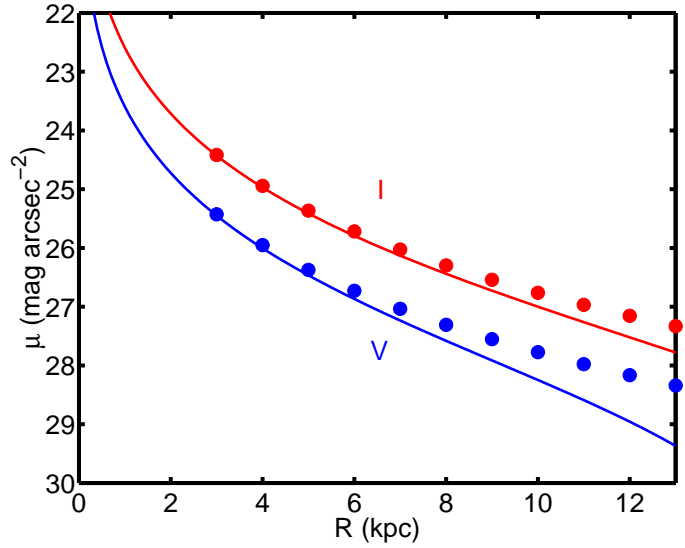}
\caption{Schematic illustration of how halo surface brightness profiles in $V$ and $I$, as inferred from surface photometry (lines) and star counts (filled circles), are predicted to deviate from each other if EBL extinction is affecting the halo colours. EBL extinction causes a higher flux deficit in the $V$-band than in $I$, thereby giving rise to anomalously red $V-I$ colours in the outskirts. These predictions assume the surface brightness of the diffuse EBL derived by Zackrisson et al. \cite{Zackrisson et al. c}, a Milky Way extinction law and a $V$-band halo opacity of $A_V=0.05$ mag. Given these parameters, surface photometry measurements of $V-I$ would indicate $V-I=1.2$ (which is at the limit of what any normal, metal-poor stellar population can achieve, based on theoretical isochrones at $Z\leq 0.008$ \cite{Marigo et al.}) at $\mu_I\approx 27$ mag arcsec$^{-2}$ and a colour that becomes progressively redder beyond that point (i.e. $V-I>1.2$).}
\label{fig1}
\end{figure}

The EBL extinction scenario gives testable predictions for how surface brightness profiles based on star counts and surface photometry should deviate from each other. If red halos really are due to EBL extinction, the surface photometry profiles should drop off faster than the star counts, whereas other explanations for the red halo colours \cite{Bergvall et al.,Zackrisson et al. b} would give the opposite effect. This drop-off is schematically illustrated in Fig.~\ref{fig1}. To implement this test, one needs to target a galaxy  for which star counts and surface photometry can both be applied across the halo isophotes where the red excess is likely to turn up. This turns out to be a non-trivial optimization problem. Star counts can only be applied for nearby galaxies, which typically have large angular sizes. Surface photometry, on the other hand, becomes increasingly difficult once the angular size of the target starts to fill the field of view of the telescope. In a recent paper \cite{Zackrisson et al. a}, we argue that the target galaxies need to be located at distances $\lesssim 16$ Mpc, provided that the star counts are carried out with Hubble Space Telescope (HST) resolution. Luckily, HST halo observations already exist for many such galaxies (e.g. obtained as part of the GHOSTS survey \cite{de Jong et al.}). To implement the proposed test, one just needs to obtain matching surface photometry data.

\section{Measuring the optical EBL}
If the red halo colours can be demonstrated to be due to EBL extinction, this could provide interesting constraints on the EBL itself. For halo opacities in the $A(g)\sim 0.01$--0.1 range (in agreement with current measurements), the surface brightness of the optical EBL would have to be at a level similar to that suggested by the Bernstein \cite{Bernstein} EBL measurements. This is considerably higher than what currently resolved galaxies can account for (e.g. \cite{Madau & Pozzetti}), implying that the origin of the optical EBL remains unknown, with potentially far-reaching implications for cosmology. An independent constraint of this type would also help break the current stalemate between the direct optical EBL measurements \cite{Bernstein}, and the EBL constraints based on gamma-ray observations of blazars (e.g. \cite{Albert et al.}) which suggest an EBL level closer to that produced by resolved galaxies. 

If this EBL extinction signature (Fig.~\ref{fig1}) is detected, the offset between the profiles can be converted into an independent measurement of the optical EBL, provided that the halo opacity can also be assessed. In principle, an EBL measurement of this type can be obtained as long as the surface brightness profile based on surface photometry drops faster than the one based on star counts, even if this does not result in abnormally red halo colours. As argued by Zackrisson et al. \cite{Zackrisson et al. c}, a situation of this type can occur for certain combinations of effective extinction curves and EBL colours.  

\begin{figure}
\centering
\includegraphics[width=.6\textwidth]{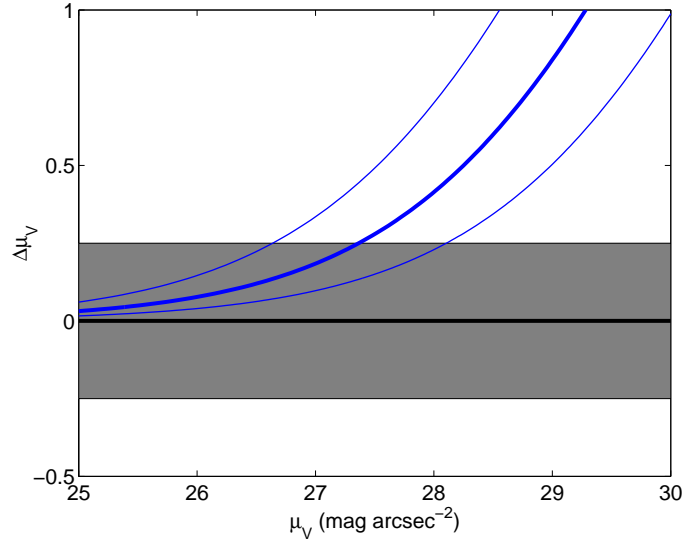}
\caption{Deviation $\Delta \mu_V$ between the surface photometry and star count profiles, as a function of the halo surface brightness $\mu_V$ based on surface photometry. For an assumed halo opacity of $A(V)=0.05$ mag, the deviation is predicted to fall on the thick blue line (with thin blue lines indicating the expected range in the case of a factor of two uncertainty on $A(V)$) for a diffuse EBL at $\mu_{\mathrm{EBL},V}=25.5$ mag arcsec$^{-2}$ -- the value derived by Zackrisson et al. (\cite{Zackrisson et al. c}), based on data from Bernstein \cite{Bernstein}. For an assumed measurement error of $\pm 0.25$ mag on $\Delta \mu_V$ (grey region, arbitrarily centered on $\Delta \mu_V=0$), a measurement of the optical extinction $A(V)$ at $\mu_V> 28$ arcsec$^{-2}$ should be able to rule out (measurement within grey region) or confirm (measurement within thin blue lines) a diffuse EBL as bright as $\mu_{\mathrm{EBL},V}=25.5$ mag arcsec$^{-2}$.}
\label{fig2}
\end{figure}

In Fig.~\ref{fig2}, we plot the difference $\Delta \mu_V$ between the star counts and surface photometry measurements as a function of the $V$-band halo surface brightness based on surface photometry. For a halo opacity of $A(V)=0.05$ mag, the predicted deviation $\Delta \mu_V$ is given by the thick blue line, with the thin blue lines indicating the limiting cases for a factor of 2 uncertainty on $A(V)$ (i.e. $^{+0.05}_{-0.025}$ mag). This prediction is based on a diffuse EBL level of  $\mu_{\mathrm{EBL},V}=25.5$ mag arcsec$^{-2}$, which corresponds to the EBL surface brightness inferred from the direct measurements \cite{Bernstein}. The adopted opacity $A(V)=0.05^{+0.05}_{-0.025}$ is broadly consistent with the Zaritsky \cite{Zaritsky} and M\'enard et al. \cite{Menard et al.} measurements of the opacity in the inner region of galaxy halos. It is also in the range required to explain red halos as due to EBL extinction \cite{Bergvall et al.,Zackrisson et al. c}. If one assumes that $\Delta \mu_V$ can be measured with an accuracy of $\pm 0.25$ mag at $\mu_V> 28$ arcsec$^{-2}$, this would provide strong evidence for or against a diffuse EBL as bright as that measured by Bernstein. If, on the other hand, the diffuse EBL is at level similar to or fainter than the EBL currently resolved into galaxies ($\mu_V\geq 27.9$ mag arcsec$^{-2}$; \cite{Zackrisson et al. c,Madau & Pozzetti}), one expects the deviation between star counts and surface photometry to be much closer to zero (within the grey regions, which indicate the assumed $\pm 0.25$ mag error on $\Delta \mu_V$).

The most difficult part of this endeavour is likely to be the measurement of the halo opacity. While a full treatment of this problem is outside the scope of this paper, we argue that the most promising approach may be to use the {\it Herschel}\footnote{www.esa.int/herschel} telescope to measure the far-IR/submm dust emission spectrum in the same region of the halo as the star counts and surface photometry measurements. To realise this in practice, however, will require quantitative consideration of a number of factors influencing the brightness and colour of the dust emission.

For the measurement of the optical EBL, most of the EBL will be attenuated by large grains in equilibrium with the radiation field of the host galaxy. In this case, the colour of the FIR/submm emission from the grains effectively constrains the grain heating rate, allowing the grain column density to be directly inferred from infrared observations. The derived grain column density can then be converted into an absorption opacity at any desired optical wavelength, provided the ratio of the grain absorption coefficients in the optical and FIR/submm is known. Detailed models of the FIR/submm dust emission spectrum can be produced using radiative transfer calculations of the type recently carried out by Popescu et al. \cite{Popescu et al.}, but the dust emission spectrum is likely to peak in the 150 -- 250 micron range, depending on the exact location within the halo and the star formation properties of the disc. This is well matched to the grasp in wavelength of the {\it Herschel/PACS} and {\it Herschel/SPIRE} instruments. 

One potential complication is that the grains in the halo could be self-shielding. If there were to be (at the wavelength where the EBL is to be measured) a population of optically thick clouds then in general one would tend to overestimate the true absorption factor of the diffusely distributed dust. At the limited angular resolution of IR facilities, the only way of recognising such clouds would be through detection of the expected very cold population of dust grains from the interior of the clouds. This would potentially show up in the form of an emission excess in the deep submm regime. For this reason it will advisable to extend the measurements deep into the submm, by also embracing the 350 and 500 micron channels of {\it Herschel/SPIRE}.

\section*{Acknowledgments}
Nils Bergvall, Roelof de Jong, G\"oran \"Ostlin, Chris Flynn and Brady Caldwell are acknowledged for fruitful collaboration on the topic of red halos. R.J. Tuffs is acknowledged for substantial input on the method of using the FIR/sub-mm dust emission spectrum to measure halo opacities. E.Z. acknowledges research grants from the Swedish National Space Board, the Swedish Research Council and the Royal Physiographical Society of Lund.

\end{document}